\documentclass[letterpaper,twocolumn,showpacs,preprintnumbers,amsmath,amssymb,floatfix,superscriptaddress]{revtex4}

\usepackage{epsfig}
\usepackage{amsmath}

\begin{document}

\title{Synchronization in a neuronal feedback loop \\ through asymmetric temporal delays}

\author{Sebastian F.~Brandt}
\affiliation{
Department of Physics,
Washington University in St.~Louis, MO 63130-4899, USA}

\author{Axel Pelster}
\affiliation{                                  
Fachbereich Physik, Universit{\"a}t Duisburg-Essen,  Lotharstra{\ss}e 1,
47048 Duisburg, Germany}      

\author{Ralf Wessel}
\affiliation{
Department of Physics,
Washington University in St.~Louis, MO 63130-4899, USA}

\date{September 14, 2007}

\begin{abstract}
We consider the effect of asymmetric temporal delays in a system of two coupled Hopfield neurons. For couplings  of opposite signs, a limit cycle emerges via a supercritical Hopf bifurcation when the sum of the delays reaches a critical value. We show that the angular frequency of the limit cycle is independent of an asymmetry in the delays. However, the delay asymmetry determines the phase difference between the periodic activities of the two components. Specifically, when the connection with negative coupling has a delay much larger than the delay for the positive coupling, the system approaches in-phase synchrony between the two components. Employing variational perturbation theory (VPT), we achieve an approximate analytical evaluation of the phase shift, in good agreement with numerical results.
\end{abstract}


\maketitle

Synchronization phenomena among coupled systems are abundant in nature \cite{Pikovsky,Strogatz}. The coupling is often not instantaneous; rather finite time delays exist.  In general, time delays can cause an otherwise stable system to oscillate  \cite{Heiden,Coleman,Hadeler} and may lead to bifurcation scenarios resulting in chaotic dynamics \cite{Wischert,Schanz}. For example, delay-induced oscillations have been reported for neural networks \cite{DelIndOsc}, genetic regulatory networks \cite{GenRegNet}, and models of population dynamics \cite{PopDyn} to name just a few.  The delays for the different coupling mechanisms in such networks do not need to be uniform, which may have an important effect on the system dynamics. For instance, it has been shown that distributed delays can stabilize a dynamical system \cite{Eurich}, and the influence of delayed inhibitory feedback has been studied \cite{Glass}.  In regard to network synchrony, the question arises under what conditions this special form of network behavior can be maintained when the temporal delays are nonuniform.
 
Asymmetric time delays in the visual pathway can be a pathological condition, as they are associated with many diseases \cite{Mojon}.  However, when feedback loops in biological systems have evolved to feature different latencies for feed-forward and feedback projections, this might provide a hint that asymmetric delays can also be beneficial to a system's functioning. In the avian visual system, the optic tectum is reciprocally coupled with the nucleus pars parvocellularis (Ipc), a subnucleus of the nucleus isthmi \cite{Luksch}. The coupled systems, tectum and Ipc, respond with synchronized oscillatory bursts to visual stimulation \cite{Marin}. Remarkably, the Ipc axons projecting to the tectum are thick and myelinated (fast action potential propagation), whereas tectal neurons projecting to the Ipc possess comparatively thin axons and are unmyelinated (slow action potential propagation) \cite{Luksch}. The Ipc-to-tectum delay may thus be as short as a fraction of a millisecond, whereas the delay for the tectum-to-Ipc projection can be expected to be of the order of tens of milliseconds. It therefore seems natural to conjecture that the asymmetry in the delays may play a functional role in the feedback system.

To explore this conjecture we investigate a model system of two coupled Hopfield neurons \cite{Hopfield} with asymmetric delays, described by the coupled first-order delay differential equations (DDE's)
\begin{eqnarray}
\frac{d u_1 (t)}{dt} &=& -u_1(t) + a_1 \tanh[u_2(t - \tau_2)] 
\, , \nonumber \\
\frac{d u_2 (t)}{dt} &=& -u_2(t) + a_2 \tanh[u_1(t - \tau_1)] \label{ddesys} 
\,.
\end{eqnarray}
Here,  $u_1$ and $u_2$ denote the voltages of the Hopfield neurons and $\tau_1$ and $\tau_2$ are the signal propagation or processing time delays, while $a_1$ and $a_2$ describe the couplings between the two neurons. 
The system of DDE's (\ref{ddesys}) has a trivial stationary point at $u_1 = u_2 = 0$, the stability of which has been analyzed in detail, e.g., in ref.~\cite{Wei}.  For $a_1 a_2 < -1$ the fixed point at the origin is asymptotically stable as long as the mean of the time delays $\tau \equiv (\tau_1 + \tau_2)/2$ does not exceed the critical value
$\tau_{0} \equiv \sin^{-1}[-2 \omega_0 / (a_1a_2)]/(2\omega_0)$,
where $\omega_0 = \sqrt{\vert a_1 a_2 \vert - 1}$.  When the sum of the delays is increased, the origin becomes unstable and a limit cycle emerges via a supercritical Hopf bifurcation at $\tau = \tau_0$.  Note that the characteristic equation for the system (\ref{ddesys}), which determines the condition for a periodic solution to exist, only depends on the sum of the two delays. A linear stability analysis can thus provide no insight toward a possible role of asymmetry in the delays.  Furthermore, standard methods for bifurcation analysis, as described, e.g., in refs.~\cite{Wischert,Redmond} are only suitable for examining the nonlinear dynamical system in the immediate neighborhood of the bifurcation. In contrast to that, in this letter we aim at obtaining results that also hold for large delays, i.e., far away from the bifurcation.
\begin{figure}[t]
  \begin{center}
    \epsfig{file=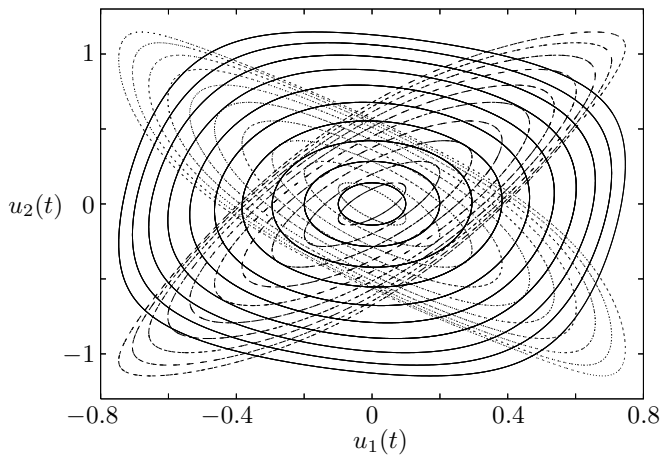,width=\columnwidth}    
    \caption{Numerical solutions to the system of DDE's (\ref{ddesys}) for the choice of parameters $a_1 = -1$ and $a_2 = 2$ and for different values of the time delays $\tau_1$, $\tau_2$ (transients not shown). Solutions for the case $\tau_1 = \tau_2$ are represented by solid lines. Dashed and dotted lines represent solutions for the cases $\tau_1 = 0$ and $\tau_2 = 0$, respectively.  For each set of lines the value of the delay parameter $\epsilon = \sqrt{\tau - \tau_0}$ increases from the innermost limit cycle ($\epsilon = 0.1$) to the outermost limit cycle ($\epsilon = 1.0$) in increments of $\Delta \epsilon = 0.1$.}
  \label{fig1}
  \end{center}
\end{figure}   
\begin{figure}[t]
  \begin{center}
    \epsfig{file=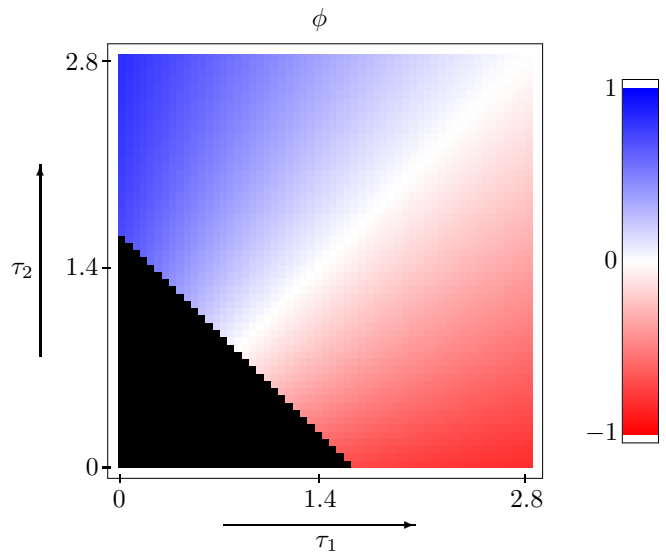,width=\columnwidth}    
    \caption{(Color) Plot of the phase shift between $u_1(t)$ and $u_2(t)$.  Numerical results for the scalar product $\phi$ as given by (\ref{phi}) are color coded for combinations of $\tau_1$ and $\tau_2$ with $0 \leq \tau_{1/2} \leq 2.8$. Red and blue indicate negative and positive values of $\phi$, respectively.  In the black region, no periodic solution exists.}
    \label{fig2}
  \end{center}
\end{figure}

We first investigate the effect of asymetric time delays through numerical simulations.  For a subsequent perturbation expansion we define the expansion parameter $\epsilon = \sqrt{\tau - \tau_0}$. Figure \ref{fig1} shows numerical solutions of the system of DDE's (\ref{ddesys}) for different values of the time delays $\tau_1$ and $\tau_2$ and for the choice of parameters $a_1 = -1$, $a_2 = 2$.  The amplitude of the limit cycle is only determined by the value of $\epsilon$ and thus remains unchanged when the temporal delays are chosen to be different.  However, we observe that the phase between the periodic activities of $u_1(t)$ and $u_2(t)$ does depend on the asymmetry of the delays.  In order to quantify this phase difference, we consider the normalized scalar product
\begin{eqnarray}
\phi = \frac{\int_{T_0}^{T_0 + T} dt \hspace{1mm} u_1(t)u_2(t)}{\left[\int_{T_0}^{T_0 + T} d t \hspace{1mm} u_1(t)u_1(t) \int_{T_0}^{T_0 + T} d t \hspace{1mm} u_2(t)u_2(t)\right]^{1/2}}\, . \label{phi}
\end{eqnarray}
Numerical results for this quantity are shown in fig.~\ref{fig2}. We find that for time delays which are equal or at least not too asymmetric the scalar product $\phi$ is approximately zero, which corresponds to a phase shift of $\pi/2$ between $u_1(t)$ and $u_2(t)$, assuming that they can be described by sinusoidal functions. However, when the delays are asymmetric, the scalar product $\phi$ becomes larger in magnitude, being negative for $\tau_1 > \tau_2$ and positive for $\tau_2 > \tau_1$.  Specifically, for $\tau_1 = 0$ the scalar product approaches unity for a growing delay $\tau_2$, corresponding to in-phase synchronization between the the two components.

We now aim at achieving an approximate analytical calculation of $\phi$.  To this end, we first derive the perturbation series for the periodic solution $\mathbf{u}(t)$ and its angular frequency $\omega$ of the system (\ref{ddesys}) by applying the  Poincar\'e-Lindstedt method \cite{MacDonald}.  Since a supercritical Hopf bifurcation occurs at $\tau = \tau_0$, we assume that the amplitude and frequency of the new periodic states are analytic in $\epsilon$ and expand them as
$
\mathbf{u}(t) = \epsilon \mathbf{U}(t) = \epsilon \left[ \mathbf{U}^{(0)}(t) + \epsilon \mathbf{U}^{(1)}(t) + \ldots \right]$, 
$\omega(\epsilon) = \omega_0 + \epsilon \omega_1 + \epsilon^2 \omega_2 + \ldots$.
Furthermore, for convenience we introduce the rescaled independent variable $\xi = \omega(\epsilon)t$ and write $\mathbf{U}(t) = \mathbf{V}(\xi)$. 
The expansion then proceeds in a way very similar to the approach in ref.~\cite{LimVPT}, where the frequency of the limit cycle is calculated perturbatively for increasing mean of time delays. However, we introduce an additional parameter $\tilde{\tau}_1$, which is defined as the $\tau_1$-value of the intersection point in the $\tau_1$-$\tau_2$ plane between the line that marks the boundary between the regions in which a periodic solution does or does not exist, and a line perpendicular to this boundary through a given point $(\tau_1,\, \tau_2)$. 
To $n$th order in $\epsilon$, we have to solve a system of differential equations of the form
\begin{figure*}[t]
  \begin{center}
    \epsfig{file=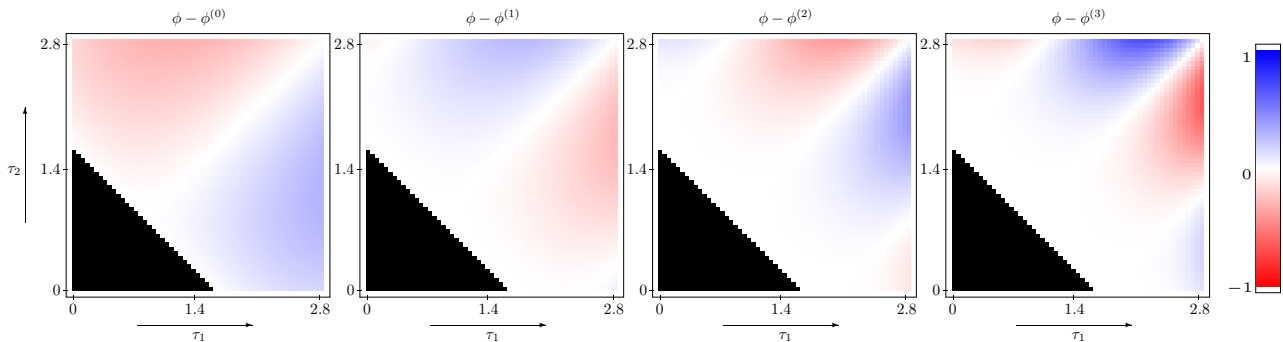,width=17cm}    
    \caption{(Color) Perturbative results for the phase shift between $u_1(t)$ and $u_2(t)$. The color-coded plots show the difference between the numerical result from fig.~\ref{fig2} and the perturbative results up to order $g^3$.}
\label{fig3}
  \end{center}
\end{figure*}
\begin{eqnarray}
\frac{d V_1^{(n)}(\xi)}{d \xi} &=& - \frac{V_1^{(n)}(\xi)}{\omega_0} \label{VnDEsys} \\ && + \hspace{1mm} \frac{a_1}{\omega_0}V_2^{(n)}[\xi - \omega_0 (2 \tau_0 - \tilde{\tau}_1)] + f_1^{(n)}(\xi)\, , \nonumber \\
\frac{d V_2^{(n)}(\xi)}{d \xi} &=& - \frac{V_2^{(n)}(\xi)}{\omega_0} + \frac{a_2}{\omega_0} V_2^{(n)}(\xi - \omega_0\tilde{\tau}_1) + f_2^{(n)}(\xi) \, , 
\nonumber
\end{eqnarray}
where the inhomogeneity $\mathbf{f}^{(n)}(\xi)$ is determined by the solutions to previous orders.  Since we require that the solution $\mathbf{V}^{(n)}(\xi)$ be periodic in $\xi$ with period $2\pi$, we can impose certain conditions on the inhomogeneity $\mathbf{f}^{(n)}(\xi)$.  Namely, we demand that $\mathbf{f}^{(n)}(\xi)$ not contain terms that would lead to non-periodic solutions for $\mathbf{V}^{(n)}(\xi)$, i.e., $\mathbf{f}^{(n)}(\xi)$ must not contain secular terms.  These conditions, which can be derived by expanding both the $n$th order limit cycle solution $\mathbf{V}^{(n)}(\xi)$ and the inhomogeneity $\mathbf{f}^{(n)}(\xi)$ into a Fourier series, read
\begin{eqnarray}
a_2 \sin(\omega_0 \tau_0) \alpha_{1,1}^{(n)}  +   \alpha_{2,1}^{(n)}\sin[\omega_0(\tau_0 - \tilde{\tau}_1)] \nonumber \\ \ +  \hspace{1mm} \beta_{2,1}^{(n)}\cos[\omega_0(\tau_0 - \tilde{\tau}_1)]  &=& 0 \, , \nonumber \\
\alpha_{2,1}^{(n)}\cos[\omega_0(\tau_0 - \tilde{\tau}_1)]  - \beta_{2,1}^{(n)}\sin[\omega_0(\tau_0 - \tilde{\tau}_1)]  \nonumber \\ - \hspace{1mm}  a_2 \sin(\omega_0 \tau_0) \beta_{1,1}^{(n)} &=& 0 \, . \label{Conds} 
\end{eqnarray}
Here $\alpha_{1/2,1}^{(n)}$ and  $\beta_{1/2,1}^{(n)}$ denote the coefficients of the cosine and sine terms in the Fourier expansion of the inhomogeneity $f_{1/2}^{(n)}(\xi)$, respectively.
Imposing these conditions on the inhomogeneity in (\ref{VnDEsys}) allows us to determine the angular frequency correction $\omega_n$ and the Fourier expansion coefficients for $\mathbf{V}^{(n - 2)}(\xi)$.  To second order in $\epsilon$ we find
\begin{eqnarray}
\omega_2 &=& -\frac{\omega_0^2}{\omega_0 \tau_0 + \cos(\omega_0 \tau_0) \sin(\omega_0 \tau_0)} \, ,  \label{om2}
\end{eqnarray}
while $\omega_1$ vanishes.  This value is identical to the one found in ref.~\cite{LimVPT} depending only on $\omega_0$ and $\tau_0$ but not on $\tau_1$ or $\tau_2$.  Since this observation holds to all orders, we thus find that the period of the oscillations is independent of any asymmetry in the time delays. Furthermore, we find that only even perturbative orders lead to nonvanishing contributions for both the angular frequency $\omega$ and the limit cycle $\mathbf{V}(\xi)$; we therefore define the new expansion parameter $g = \epsilon^2$. Denoting the expansion to order $g^N$ of the quantity (\ref{phi}) by $\phi^{(N)}$, we find
\begin{eqnarray}
\phi^{(1)} &=& \frac{\cos[\omega_0(2 \tau_0 - \tilde{\tau}_1)] - \omega_0 \sin[\omega_0(2 \tau_0 - \tilde{\tau}_1)]}{\textrm{sign}(a_1)\sqrt{1 + \omega_0^2}}  \label{phi1} \\
&& \hspace{-14mm} + g
\frac{2(\tau_0 \hspace{-0.5mm} -  \hspace{-0.3mm} \tilde{\tau}_1)\omega_0^2 \left\{\sin[\omega_0(2\tau_0 \hspace{-0.5mm} -  \hspace{-0.3mm} \tilde{\tau}_1)] \hspace{-0.7mm} +  \hspace{-0.3mm} \omega_0 \cos[\omega_0(2\tau_0 \hspace{-0.5mm} -  \hspace{-0.3mm} \tilde{\tau}_1)]\right\}^3}{ \textrm{sign}(a_1) [\sin(2\omega_0 \tau_0) \hspace{-0.5mm} + 2 \omega_0 \tau_0](1 + \omega_0^2)^{3/2}}. \nonumber
\end{eqnarray}
Focussing on the choice of parameters $a_1 = -1$, $a_2 = 2$, which leads to $\omega_0 = 1$, $\tau_0 = \pi / 4$, we can determine the expansion coefficients for $\phi^{(N)}$ up to the third order.  Figure~\ref{fig3} shows a comparison of our perturbative results and the numerical result from fig.~\ref{fig2}.  For small time delays, the accuracy of the results from the perturbation expansion is good and improves with increasing order.  However, as $g$ increases, the perturbative results cease to converge and no longer provide an acceptable approximation.  As is typical for perturbative methods, our approach has yielded a divergent series.  In order to improve the quality of our results, we now perform a resummation of the perturbative expansion employing variational perturbation theory (VPT).
\begin{figure*}[t]
  \begin{center}
    \epsfig{file=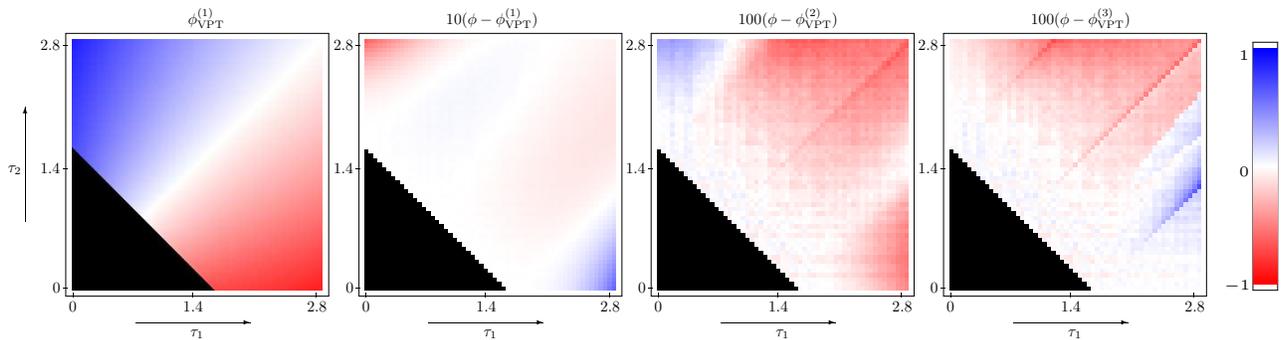,width=17cm}    
    \caption{(Color) VPT results for the phase shift between $u_1(t)$ and $u_2(t)$. The first color-coded plot shows the first-order-VPT result as given by (\ref{VPT1}).  The three other plots show the difference between the numerical result and the results from the first three orders in VPT.  For clarity, this difference has been augmented by a factor of $10$ and $100$ in the results for the first and for both the second and third order, respectively}
\label{fig4}
  \end{center}
\end{figure*}

VPT is a nonperturbative approximation scheme based on a variational approach due to Feynman and Kleinert \cite{Feynman2}, which has been systematically developed over the last few years, establishing its applicability in various fields of physics \cite{LimVPT,Kleinertsys,PathInt4,VerenaBuch,Festschrift,JankeC1}.  VPT permits the evaluation of a divergent series of the form $f^{(N)}(g) = \sum_{n = 0}^N a_n g^n$
and yields a strong-coupling expansion of the generic form $f(g) = g^{p/q}\sum_{m = 0}^M b_m g^{-2m/q}$.
Here, $p$ and $q$ are real growth parameters characterizing the strong-coupling behavior.  The convergence of the series after resummation is exponentially fast and uniform with respect to other system parameters  such as temperature,  coupling constants, spatial dimensions, etc.\ \cite{VPTApp}.  

In order to perform the resummation, one introduces a variational parameter $K$ for the perturbation series according to Kleinert's square-root trick \cite{PathInt4}. The series is thus transformed to the expression
\begin{eqnarray}
f^{(N)}(g, K) &=&  \label{fNgK} \\
&& \hspace{-12mm} \sum_{n = 0}^{N} a_n g^n K^{p - nq} \sum_{k = 0}^{N - n} \binom{(p - nq)/2}{k} 
\left( \frac{1}{K^2} - 1 \right)^k \,,  \nonumber
\end{eqnarray}
derived in detail in ref.~\cite{LimVPT}.
The influence of the variational parameter $K$ is then optimized according to the principle of minimal sensitivity \cite{Stevenson}; i.e., the optimized value $K^{(N)}$ is determined by solving for the roots of the first or higher derivatives of $f^{(N)}(g,K)$ with respect to $K$. The $N$th order VPT approximation is then obtained by evaluating (\ref{fNgK}) at this optimized value: $f^{(N)}_{\rm VPT}(g) = f^{(N)}(g, K^{(N)})$.  This variational result generally holds for all values of the coupling constant $g$.  Furthermore, by considering the limit of large $g$, it allows the extraction of the strong-coupling coefficients $b_m$.

In our case of the perturbation series for $\phi$, the values of the growth parameters $p$ and $q$ turn out to be the same as those that we determined in ref.~\cite{LimVPT} for the angular frequency, namely $p = -2$, $q = 2$.  Our first-order result after resummation then reads
\begin{eqnarray}
\phi^{(1)}_{\rm VPT}(g) &=& \label{VPT1} \frac{(2 + \pi)(1 - 2 \cos \tilde{\tau}_1 \sin \tilde{\tau}_1)}{\sqrt{2}} \\ && \hspace{-20mm} \times [(2 \hspace{-0.5mm}+\hspace{-0.5mm} \pi)( \cos \tilde{\tau}_1 \hspace{-0.5mm} -  \sin \tilde{\tau}_1) +  g(\pi - 4 \tilde{\tau}_1) ]( \cos \tilde{\tau}_1 \hspace{-0.5mm} +  \sin \tilde{\tau}_1)]^{-1}\hspace{-0.5mm}. \nonumber
\end{eqnarray}
The first color-coded plot in fig.~\ref{fig4} shows a graphical representation of this result.  The agreement with the numerical result from fig.~\ref{fig2} is excellent.  While the second VPT order provides a significant improvement when compared with the first order result, third order results are slightly superior to those of second order.

In conclusion, our investigation of a neuronal model system shows that asymmetric temporal delays can control the phase in a feedback loop and lead to synchronous oscillations. Specifically, in-phase and anti-phase synchrony arises when the delays are maximally asymmetric. Furthermore, after a variational resummation of the perturbation series for $\phi$, we have a very accurate approximate result for this quantity even in low orders and throughout the full $\tau_1$-$\tau_2$ plane.

\acknowledgments
We wish to acknowledge assistance from Michael Schanz in solving the system of DDE's (\ref{ddesys}) numerically.  
We thank John Clark and Hagen Kleinert for critical reading of the manuscript.
This work was supported in part by NIH-EY 15678. One of us, S.~F.~B., acknowledges support from a Grant-In-Aid of Research from Sigma Xi, The Scientific Research Society.


\end{document}